\documentclass{aa}
\usepackage{txfonts}
\usepackage{graphicx}
\usepackage{rotating}

\begin{document}

\newcommand{\inthms}[3]{$#1^{\rm h}#2^{\rm m}#3^{\rm s}$}
\newcommand{\dechms}[4]{$#1^{\rm h}#2^{\rm m}#3\mbox{$^{\rm
s}\mskip-7.6mu.\,$}#4$}
\newcommand{\intdms}[3]{$#1^{\circ}#2'#3''$}
\newcommand{\decdms}[4]{$#1^{\circ}#2'#3\mbox{$''\mskip-7.6mu.\,$}#4$}
\newcommand{\tmb}{\mbox{T$_{\rm mb}$}}
\newcommand{\OI}{\mbox{O\,{\sc i}}}
\newcommand{\dtco}{D$_{2}$CO}
\newcommand{\hdco}{HDCO}
\newcommand{\htco}{H$_{2}$CO}
\newcommand{\httco}{H$_{2}^{13}$CO}
\newcommand{\kmps}{km s$^{-1}$}
\newcommand{\APer}{$\alpha$\,Per}
\newcommand{\Msun}{\,M$_{\odot}$}
\newcommand{\BD}{brown dwarf}
\newcommand{\BDs}{brown dwarfs}
\newcommand{\CMD}{colour-magnitude diagram}
\newcommand{\CCD}{colour-colour diagram}
\newcommand{\ra}{\rightarrow}

\hyphenation{Fe-bru-ary Gra-na-da mo-le-cu-le mo-le-cu-les}


\title{A near-infrared survey for new low-mass members in $\alpha$\,Per
}

\author{N. Lodieu \inst{1,2} \thanks{Visiting Astronomer, German-Spanish 
        Astronomical Centre, Calar Alto,
        operated by the Max-Planck-Institute for Astronomy, Heidelberg, jointly
        with the Spanish National Commission for Astronomy.}
        \and
        M. J. McCaughrean \inst{2,3}
        \and
	D. Barrado y Navascu\'es \inst{4}
        \and
	J. Bouvier \inst{5}
        \and
	J. R. Stauffer \inst{6}
}

\offprints{N. Lodieu}

\institute{University of Leicester, Department of Physics \& Astronomy,
           University Road, Leicester LE1 7RH, UK \\
           \email{nl41@star.le.ac.uk}
           \and
           Astrophysikalisches Institut Potsdam, An der Sternwarte 16, 14482 Potsdam, 
           Germany
           \and 
           University of Exeter, School of Physics, Stocker Road,
           Exeter EX4 4QL, UK \\
           \email{mjm@aip.de, mjm@astro.ex.ac.uk}
           \and
	   Laboratorio de Astrof\'{\i}sica Espacial y F\'{\i}sica Fundamental, INTA, 
           P.O. Box 50727, 28080 Madrid, Spain \\
	   \email{barrado@laeff.esa.es} 
	   \and
	   Laboratoire d'Astrophysique, Observatoire de Grenoble,
	   BP 53, 38041 Grenoble C\'edex 09, France \\
	   \email{Jerome.Bouvier@obs.ujf-grenoble.fr}
	   \and
	   SIRTF Science Center, California Institute of Technology, 
           Pasadena, CA\,91125, USA \\
           \email{stauffer@ipac.caltech.edu}
}

\date{Received {\today} /Accepted }

\titlerunning{A near-infrared survey for new low-mass members in $\alpha$\,Per}
\authorrunning{Lodieu et al.}

\abstract{
We present a near-infrared ($K^{\prime}$-band) survey of 0.7 square
degree area in the $\alpha$\,Persei open cluster (age\,=\,90\,Myr,
distance\,=\,182\,pc) carried out with the 
Omega-Prime camera on the Calar Alto 3.5-m telescope.
Combining optical data ($R_{c}$ and $I_{c}$) obtained with the KPNO/MOSA detector
and presented in Stauffer et al.\ (\cite{stauffer99}) with the $K^{\prime}$
observations, a sample of new candidate members has been extracted from
the optical-infrared colour-magnitude diagram.
The location of these candidates in the colour-colour diagram
suggests that two-thirds of them are actually reddened background
giants. About 20 new candidate members with masses between
0.3 and 0.04\Msun{} are added to the $\sim$\,400 known \APer{} cluster members.
If they are indeed \APer{} members, four of the new candidates
would be brown dwarfs.
We discuss the advantages and drawbacks of the near-infrared survey
as compared to the optical selection method.
We also describe the outcome of optical spectroscopy obtained with the
Twin spectrograph on the Calar Alto 3.5-m telescope for about 30
candidates, including selected members from the optical sample
presented in Barrado y Navascu\'es et al.\ \cite{barrado02} and from
our joint optical/infrared 
catalogue.  These results argue in favour of the optical selection method 
for this particular cluster.
\keywords{Stars: low-mass, brown dwarfs -- Techniques: photometry -- Techniques: spectroscopy -- 
Infrared: stars -- Hertzsprung--Russell (HR) and C-M diagrams -- open clusters and associations: individual
($\alpha$\,Persei)} 
}

\maketitle
%

%
%
%
\section{Introduction}

\begin{figure}[htbp]
\begin{center}
\includegraphics[width=1.00\columnwidth, angle=0]{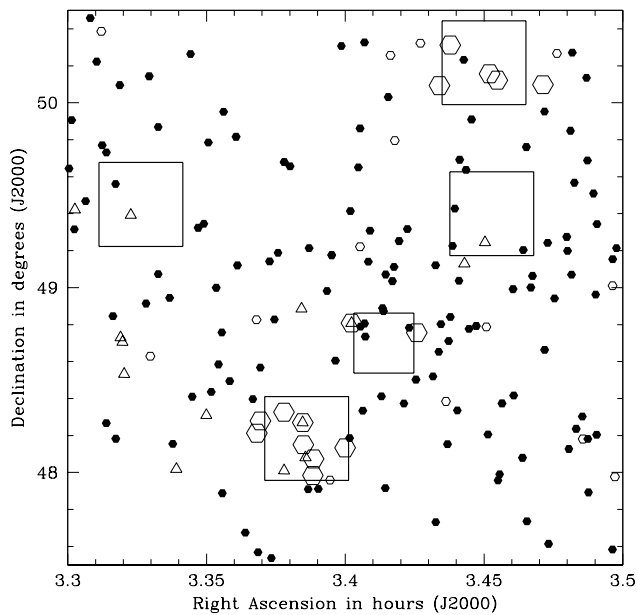}
\caption{
Location of the known \APer{} members as of 2002 around the area
covered with the Omega-Prime camera.
Filled and open circles represent the probable and
possible members, respectively, based on multi-colour
photometric surveys for low-mass members in \APer{}.
Open triangles are spectroscopically confirmed cluster
members from S99\@.
Large open hexagons are new infrared-selected candidates
described in this paper. {\bf{The five Omega-Prime fields-of-view
are plotted as black squares}}.
}
\label{APer_radec}
\end{center}
\end{figure}

The bright star Alpha Persei is surrounded by a loose open cluster covering
several square degrees, known as the \APer{} cluster.
Initial cluster member
selections carried out by Heckmann et al.\ (\cite{heckmann56}),
Stauffer et al.\ (\cite{stauffer85}, \cite{stauffer89}),
and Prosser (\cite{prosser92}, \cite{prosser94}) using
proper motions, optical photometry, and radial velocity
measurements revealed more than 300 cluster members (Fig.\ \ref{APer_radec}).
They are catalogued with names starting with HE and AP
for the bright and faint components, respectively.
The cluster was also extensively studied in X-rays with the
ROSAT satellite, confirming the membership of many photometric
candidates (Randich et al.\ \cite{randich96}; Prosser et al.\ \cite{prosser96}) 
and the identification of new cluster members
(Prosser \& Randich \cite{prosser98a}; Prosser et al.\ \cite{prosser98b};
Randich et al.\ \cite{randich98}).
Later, Basri \& Mart\'{\i}n (\cite{basri99})
used near-infrared photometry and spectroscopy to assess
membership of candidates selected by Prosser (\cite{prosser94}).
They applied the lithium depletion test (Rebolo et al.\ \cite{rebolo92})
to a handful of possible members and
inferred an age for \APer{} of between 65 and 75\,Myr,
clearly rejecting three candidates and confirming the membership
of two of them; the remainder were classified as probable
members but lacked kinematic information.
Stauffer et al.\ (\cite{stauffer99}; hereafter S99) applied the
same lithium test to a larger sample and derived an age
of 90\,Myr for the cluster.
S99 also added about 35 new candidate cluster members based on
a combination of photometric and spectroscopic criteria.
Among those new members, two objects with spectral types later than M7
were classified as brown dwarfs.
Barrado y Navascu\'es et al.\ (\cite{barrado02}; ByN02)
identified 54 new candidate members based on additional optical and
infrared imaging, with 11 of these
objects classified as candidate brown dwarf members.
Based on the probable member candidates, luminosity and mass functions
were derived from 0.3\,M$_{\odot}$ down to 0.035\,M$_{\odot}$
yielding a power law index $\alpha$ equal to 0.59\,$\pm$\,0.05
(when expressed as the mass spectrum dN/dM $\propto$ M$^{-\alpha}$),
a result consistent with that obtained by different surveys
of the Pleiades (Meusinger et al.\ \cite{meusinger96};
Mart\'{\i}n et al.\ \cite{martin98};
Bouvier et al.\ \cite{bouvier98}; Hambly et al.\ \cite{hambly99};
Dobbie et al.\ \cite{dobbie02}; Tej et al.\ \cite{tej02};
Moraux et al.\ \cite{moraux03}).
Barrado y Navascu\'es et al.\ (\cite{barrado04b}) have recently re-evaluated 
the age of \APer{} (as well as the Pleiades and IC2391), yielding
85\,Myr, consistent within the errors with the S99 determination
of 90\,Myr.
Finally, Deacon \& Hambly (\cite{deacon04}) used a pre-release of
the northern hemisphere data from the SuperCOSMOS Sky Survey 
(Hambly et al.\ \cite{hambly01})
to identify new candidate proper motion members of \APer{}
down to about $R$\,=\,18 mag (corresponding to a mass of $\sim$\,0.15\Msun{}). 
They also concluded that the large majority of photometrically-selected
candidates in \APer{} and in the Pleiades have high membership
probabilities.

%
%
%

As low-mass stars and brown
dwarfs emit most of their flux at 1\,$\mu$m and beyond,
near-infrared observations  should be well-suited to increase
the number of cluster members and, therefore,
provide a better knowledge of the cluster mass function,
particularly in the substellar regime.
Near-infrared (1--2.5\,$\mu$m) surveys have proven an efficient
way to find low-mass members
and brown dwarfs in young star-forming regions, including
the Trapezium Cluster (McCaughrean et al.\ \cite{mjm95};
Hillenbrand \& Carpenter \cite{hillenbrand00};
Lucas \& Roche \cite{lucas00};
Muench et al.\ \cite{muench01}),
Chamaeleon\,I (Comer\'on \cite{comeron00}),
$\rho$\,Ophiuchus (Rieke \& Rieke \cite{rieke90};
Preibisch et al.\ \cite{preibisch01}),
IC\,348 (Luhman et al.\ \cite{luhman03}),
Taurus-Auriga (Brice\~no et al.\ \cite{briceno02}), and
Serpens (Kaas \cite{kaas99}), because 
low-mass stars and substellar objects are much brighter
when younger. In addition, intracluster extinction due to
dust hampers the detection of young objects
at optical wavelengths.

Somewhat older open clusters have been surveyed extensively
to probe the substellar regime,
using $R$--$I$ and $I$--$z$ optical colours
as a powerful discriminant to select low-mass and
brown dwarf member candidates over large areas in 
the Pleiades (Bouvier et al.\ \cite{bouvier98};
Stauffer et al.\ \cite{stauffer98};
Zapatero Osorio et al.\ \cite{zapatero99};
Dobbie et al.\ \cite{dobbie02};
Moraux et al.\ \cite{moraux03}),
$\sigma$\,Orionis (B\'ejar et al.\ \cite{bejar99}),
$\lambda$\,Orionis (Barrado y Navascu\'es et al.\ \cite{barrado04a}),
Praesepe (Pinfield et al.\ \cite{pinfield97}),
IC\,4665 (Mart\'{\i}n \& Montes \cite{martin97}),
IC\,2391 (Barrado y Navascu\'es et al.\ \cite{barrado99}, \cite{barrado01b}; \cite{barrado04b}),
NGC2547 (Jeffries et al.\ \cite{jeffries04}),
M35 (Barrado y Navascu\'es et al.\ \cite{barrado01a}), and
the \APer{} cluster (S99; ByN02).
By extending the colour baseline,
near-infrared photometry of individual candidates
can be used to weed out contaminating foreground and 
background objects (Hodgkin et al.\ \cite{hodgkin99};
ByN02), and to narrow down
the number of possible cluster members.
Optical spectroscopy of the H$\alpha$ emission
line at 6563\,\AA{}, lithium absorption at 6708\,\AA{},
as well as Na{\small I} and K{\small I} gravity-sensitive doublets
at 7665/7699\,\AA{} and 8183/8195\,\AA{} 
are key features to assess the membership of low-mass
candidates (Stauffer et al.\ \cite{stauffer95}; 
Zapatero Osorio \cite{zapatero96}).
For example, the Li 6708\,\AA{} absorption
has been used to assess the substellar status of brown dwarf 
candidates (e.\@g.\@ in $\sigma$\,Ori; Zapatero Osorio et al.\ \cite{zapatero02}
and $\lambda$\,Ori; Barrado y Navascu\'es et al.\ \cite{barrado04a}),
and to locate the lithium depletion
boundary in various open clusters, including the
the Pleiades (Stauffer et al.\ \cite{stauffer98}),
$\alpha$\,Per (S99; Basri \& Mart\'{\i}n \cite{basri99}),
IC\,2391 (Barrado y Navascu\'es et al.\ \cite{barrado99}, \cite{barrado04b}),
and NGC\,2547 (Oliveira et al.\ \cite{oliveira03};
Jeffries et al.\ \cite{jeffries03}).

In this paper, we present a wide-field near-infrared
imaging survey (broad-band $K^{\prime}$ at 2.1$\mu$m; 
Wainscoat \& Cowie \cite{wainscoat92}) of a 0.70 square
degree area in the $\alpha$\,Persei cluster aimed at finding new
member candidates, including lower-mass brown dwarf candidates than those
identified by ByN02,
and at improving the knowledge of the luminosity and mass
functions down into the substellar regime. 
By combining these new infrared data with the existing
optical data from the KPNO/MOSA survey detailed in S99,
we have extracted a list of $\sim$\,100 new candidates,
including new brown dwarf candidates. We assume an age of 90\,Myr 
(S99) and a distance of 182\,$\pm$\,8\,pc for the cluster.
Throughout, the latter is a compromise between
the HIPPARCOS (190.5$^{+7.2}_{-6.7}$\,pc; Robichon et al.\ \cite{robichon99})
and main-sequence fitting distances 
(176.2\,$\pm$\,5.0\,pc; Pinsonneault et al.\ \cite{pinsonneault98}).
In \S\ref{WFNIR}, we discuss the membership probability
of the new candidates based on their location in 
colour-magnitude and colour-colour diagrams.
In \S\ref{specfollow}, we present the spectral classification
of nine candidates selected from the joint
optical/near-infrared catalogues. Additional spectroscopy for
{\bf{optically-selected candidates (24 probable members, 1 possible,
and 4 non-members}}) extracted by ByN02, as well as for four new 
infrared-selected candidates, is described
along with their H$\alpha$ and gravity measurements.

%
%
\begin{table*}[t]
\begin{center}
 \caption[]{Central coordinates (J2000) of the
            35.4${\arcmin}$\,$\times$35.4${\arcmin}$
            optical fields-of-view from S99 observed in the near-infrared, along with the
            observing date, exposure times, area covered, and limiting $K^{\prime}$ magnitude.
            The $K^{\prime}$ magnitude limit is the
            completeness limit defined as the point where the histogram
            of the number of stars per magnitude bin stops increasing.
            The last column provides only the range in zero points for each
            field of view, as there are 16 mosaics for the A, C, E, K, and L
            fields and 9 mosaics for the C field. Only a fraction (22--46\,\%)
	    of each MOSA field was covered in the near-infrared.
            }
 \label{log_obs}
 \begin{tabular}{ccccccc}
 \hline
 \hline
\noalign{\smallskip}
Field &   R.A.    &        Dec   &   Obs. Date &  Exp. Time & Area & $K^{\prime}_{\rm complete}$ \cr
\noalign{\smallskip}
 \hline
\noalign{\smallskip}
A  & 03 27 10 & +49 24  & 04--06\,Dec\,98 & 20\,min & 580 arcmin$^{2}$ & 17.5 \cr
C  & 03 24 50 & +48 42  & 06\,Dec\,98     & 20\,min & 280 arcmin$^{2}$ & 17.5 \cr
E  & 03 19 35 & +49 27  & 14\,Dec\,00     & 20\,min & 580 arcmin$^{2}$ & 15.5 \cr
K  & 03 27 00 & +50 13  & 11--12\,Dec\,00 & 20\,min & 580 arcmin$^{2}$ & 17.5 \cr
L  & 03 23 10 & +48 11  & 12--13\,Dec\,00 & 20\,min & 580 arcmin$^{2}$ & 17.5 \cr
\noalign{\smallskip}
 \hline
 \end{tabular}
\end{center}
\end{table*}

%
%
\section{Wide-field near-infrared observations}
\label{WFNIR}
\subsection{Observations}
\label{wf_obs}

The observations presented in this paper were carried 
out with the wide-field near-infrared camera Omega-Prime mounted
on the Calar Alto 3.5-m telescope (Table \ref{log_obs})
and cover a total area of 0.70 square degrees in $\alpha$ Per.
Omega-Prime has a 1024\,$\times$\,1024 pixel Rockwell HAWAII HgCdTe 
detector with a spatial scale of 0.396${\arcsec}$/pixel, yielding a
6.7${\arcmin}$\,$\times$\,6.7${\arcmin}$ field-of-view
(Bizenberger et al.\ \cite{bizenberger98}).
The first set of data was obtained on 4--6 December 1998
under good seeing ($\sim$\,1${\arcsec}$) but non-photometric 
conditions. Mosaics of 3\,$\times$\,3 and 4\,$\times$\,4 Omega-Prime
fields-of-view were obtained for the A and C
optical fields of S99, respectively. The second set of data
was obtained on 11--14 December 2000.
The seeing was variable
between 1 and 1.4 arcsec with non-photometric conditions
during the first three nights, and
observations were made of two 4\,$\times$\,4 mosaics
in the K and L fields. The last optical field,
the E field, was observed on 14 December 200
under poorer weather conditions and is, therefore, shallower
than the other fields.

Twenty exposures of 1 min each dithered by
less than 40${\arcsec}$ around the centre of each field were obtained
in order to subtract the sky and keep good signal-to-noise over
the largest area possible.
Differential (lights on\,--\,off) dome flat fields were taken
before and after each night of observations,
as well as dark frames with exposure times similar to 
those of our targets.

As shown in Table \ref{log_obs}, the area covered in each field
with Omega-Prime was 280 and 580 square arcmin, i.e.\ between
22 and 46\,\% of the 1250 square arcmin covered in each field
in the optical observations with the MOSA detector on the KPNO
4-m telescope (see S99 and ByN02 for more details).

The optical fields-of-view listed
in Table \ref{log_obs} were observed
in $R_{c}$ and $I_{c}$ bands with the MOSA detector mounted
on the Kitt Peak 4-m telescope (see S99
and ByN02 for more details), and
followed-up in the near-infrared using Omega-Prime.
The mosaics observed in the infrared do not match completely the
MOSA survey due to the larger optical field-of-view.

%
%
\subsection{Data reduction}
\label{wf_data_red}

The observing procedure and the data reduction method were
identical for both sets of infrared data.
A median flat-field frame was computed each night using the
tungsten-illuminated dome flats with the lights on and off.
Each on-source frame was sky subtracted
using the median of the remaining dithered images and 
divided by the flat field.
Then, the individual images were aligned and stacked to create a
20 min-exposure mosaic.

Aperture photometry was carried out with a small aperture with the radius
of the order of the FWHM for each detection (task daofind in IRAF). 
The flux of a few relatively bright
and isolated stars was measured for different aperture sizes (from
1 to 4 times the FWHM) to compute the aperture correction.
Zero point corrections to the photometry were obtained by comparison
to 2MASS $K_{s}$ magnitudes for stars in common with the
Omega-Prime data for one or more subsections within
each Omega-Prime mosaic image (typically 20 to 40 stars in each).
No colour terms were applied to the photometry as no equation are
available to convert $K'$ magnitudes into the standard system.
The zero points differ for all individual
Omega-Prime fields-of-view and have been applied to each field.
The uncertainty on the zero-points is about 0.05 mag.
The completeness limit of each field is also listed in Table \ref{log_obs}.

These limits are defined as the magnitude where
the histogram of the number of stars per bin of magnitude stops increasing,
a valid procedure as the great majority of sources are background
field stars which should continue to increase in numbers towards
fainter magnitudes. Galactic models towards the \APer{}
cluster line of sight predict a rising luminosity function
up to $K$\,$\sim$\,21.0 mag (Annie Robin, personal communication). Therefore, we
believe that the magnitude at which the observed luminosity stops increasing
is our completeness limit  and not an intrinsic feature
of the field luminosity function, as it happens three magnitudes
brighter than the model predictions.

%
\subsection{Optical-infrared catalogue}
\label{opt_ir_cat}

\begin{figure}[!h]
\begin{center}
\includegraphics[width=\columnwidth, angle=0]{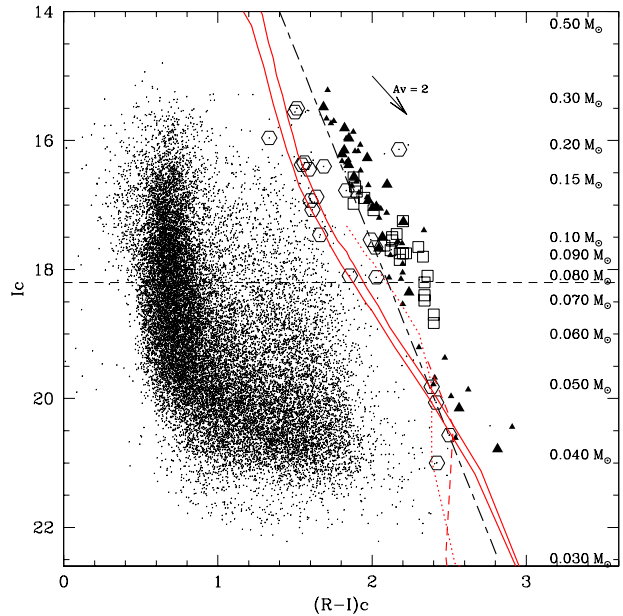}
\caption{
Colour-magnitude (($R$--$I$)$_{c}$,$I_{c}$) for all optical
detections within the 0.70 square degree area surveyed in the
near-infrared with the Omega-Prime camera on the Calar Alto
3.5-m telescope. Large filled
triangles are probable candidates from ByN02
recovered in our optical-infrared selection whereas small filled
triangles were not recovered {\bf{since the total area covered
in the near-infrared was only a fraction of that covered in the
optical}}. Open hexagons
indicate the new candidates to the right of the NextGen
100\,Myr isochrones (Baraffe et al.\ \cite{baraffe98})
in the optical-to-infrared colour-magnitude diagram.
Open squares are spectroscopically confirmed cluster members from S99\@.
Overplotted are the NextGen
isochrones for 50 and 100\,Myr (solid lines), 
the Cond (dotted line; Baraffe et al.\ \cite{baraffe03}), and
Dusty (dashed line; Chabrier et al.\ \cite{chabrier00}) 100\,Myr isochrones,
assuming a distance of 182\,pc.
The fiducial cluster main-sequence used by ByN02 to select
members in the optical is drawn as a short dash-long dash line.
The horizontal dashed line at $I_{c}$\,=\,18.2 mag
indicates the stellar/substellar boundary at 0.075\Msun{}, assuming
a distance of 182\,pc and an age of 90\,Myr for $\alpha$\,Per (S99).
The mass scale is indicated on the right side of the plot for the
assumed age and distance using the NextGen models
(Baraffe et al.\ \cite{baraffe98}).
{\bf{We have assumed a reddening value of A$_{V}$\,=\,0.30
(A$_{I}$\,=\,0.179; A$_{K}$\,=\,0.034; 
E($I_{c}$--$K$)\,=\,0.145; E($R$--$I_{c}$)\,=\,0.067)
throughout the paper, as in ByN02.}}
}
\label{phot_rii}
\end{center}
\end{figure}

Prior to the infrared survey presented in
this paper, the wide-field optical survey carried
out with the KPNO/MOSA detector (ByN02) had
revealed $\sim$\,100 member candidates in the \APer{} cluster as extracted
from the (($R$--$I$)$_{c}$,$I_{c}$) colour-magnitude diagram (Fig.\ \ref{phot_rii}).
As discussed extensively in that paper, near-infrared
follow-up of the candidates in $J$ and $K$ had allowed us to reject
some objects based on their location in optical/infrared
colour-magnitude diagrams. Three subsamples were defined:
54 objects were classified as probable members; 
12 as possible member candidates; and the remainder (26 objects)
excluded as cluster members, implying a contamination of 25--40\,\%.

Initial astrometry for our $K^{\prime}$ survey was done by
matching the observed coordinates (x, y) and the (RA, dec)
coordinates from the USNO-A2 catalogue for three relatively 
bright stars available in one Omega-Prime field-of-view.
After this registration, a higher number of stars (50 to 100)
was used to map the field distortion in Omega-Prime,
yielding RMS errors on the order of 0.1--0.2 arcsec.

%
%
%

We then merged our near-infrared ($K^{\prime}$) catalogue with the
optical ($R_{c}$ and $I_{c}$) catalogue from MOSA, using matching radius of four
times the RMS dispersion values of of 0.243${\arcsec}$ and
0.193${\arcsec}$ in $\alpha$, $\delta$, respectively.
The final catalogue provides optical ($R_{c}$ and $I_{c}$)
and infrared ($K'$) photometry for
more than 22,000 stars in a 0.70 square degree area in the \APer{} cluster.
{\bf{Only a fraction of each MOSA optical field was covered by the
near-infrared observations and therefore only part of the full MOSA
catalogue has been supplemented with infrared data in this study.}}
Table \ref{IR_detection}, available via the CDS web page,
contains 22,129 entries with the following information.
Columns 2 and 3 list the optical Right Ascension and declination (in J2000);
Column 4 and 5 list the near-infrared Right Ascension and
declination (in J2000); Columns 6, 7, and 8 give the
$R_{c}$, and $I_{c}$, and $K^{\prime}$ magnitudes along with their
associated errors. Column 9 provide the name for each source
according to IAU convention.

%
\subsection{Colour-magnitude diagram} 
\label{cmd}

\begin{figure}[htbp]
\begin{center}
\includegraphics[width=\columnwidth, angle=0]{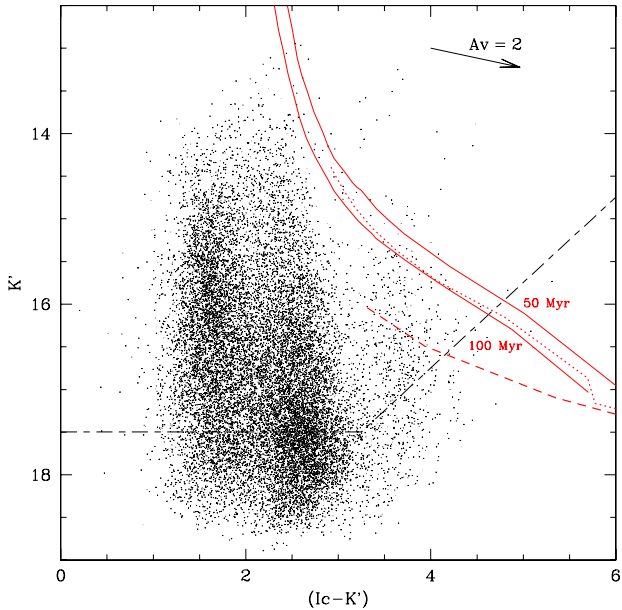}
\caption{
Colour-magnitude ($I_{c}$--$K^{\prime}$,$K^{\prime}$).
More than 22,000 detections are plotted as black dots.
Overplotted are the NextGen isochrones from Baraffe et al.\ (\cite{baraffe98})
for 50 and 100 Myr (solid lines), 
the Cond (dotted line; Baraffe et al.\ \cite{baraffe03}) and 
Dusty (dashed line; Chabrier et al.\ \cite{chabrier00}) 100\,Myr isochrones,
assuming a distance of 182\,pc for \APer{}.
Our completeness limit is represented by the short dash-long dash line.
}
\label{phot_kik}
\end{center}
\end{figure}

The main purpose of the $K^{\prime}$ survey was
to identify new cluster members, including lower mass brown dwarfs,
when combined with the optical
observations. Assuming an age of 90\,Myr derived by the lithium
depletion method (S99) and a distance of 182\,pc,
a 20\,M$_{\rm Jup}$ brown dwarf in the \APer{} cluster has an ($I$-$K$) colour
greater than 6 and a $K$ magnitude of about 17.0,
according to Baraffe et al.\ (\cite{baraffe98}) models.
Therefore, the completeness limit of the $K^{\prime}$ 
survey is deep enough to probe
the substellar regime down to 20\,M$_{\rm Jup}$. 
The optical-to-infrared colour-magnitude diagram
($I_{c}$-$K^{\prime}$,$K^{\prime}$) is shown
in Fig.\ \ref{phot_kik}. All sources are drawn as black dots. The solid
lines are the NextGen isochrones from Baraffe et al.\ (\cite{baraffe98})
for 50 and 100\,Myr, respectively, bracketing the age of the cluster.
The dotted and dashed lines are the Cond and Dusty models from
Baraffe et al.\ (\cite{baraffe03}) and Chabrier et al.\ (\cite{chabrier00}), respectively.
From Fig.\ \ref{phot_kik}, it is evident that the $K$-band data constitute
a limit to the optical survey for the two sequences of field dwarfs.
On the contrary, the optical survey represents a limit to our
near-infrared survey to extract new cluster members in the \APer{} cluster
due to the cut-off line seen from $I$--$K$,$K$\,$\sim$\,3.5,19 mag to
$I$--$K$,$K$\,$\sim$\,6,16.5 mag. A similar conclusion
can be drawn from Fig.\ \ref{phot_rii}, suggesting that our infrared
survey is limited by the $R_{c}$-band observations
($R_{c}$\,$\leq$\,21.9, ($R$--$I$)$_{c}$\,$\sim$\,2.4).
Deeper optical observations are required to check
whether or not lower mass brown dwarfs can be extracted
from the (($R$--$I$)$_{c}$,$I_{c}$) colour-magnitude diagram shown
in Fig.\ \ref{phot_rii}.

%
%
%

Three sequences are clearly visible in the ($K^{\prime}$,$I_{c}$--$K^{\prime}$)
colour-magnitude diagram (Fig.\ \ref{phot_kik}). The blue side
of the diagram (with $I$--$K$\,$<$\,2) is populated on the bright
end with F and G dwarfs from the Galactic thin disk, while fainter
objects of this colour are primarily F and G dwarfs from the
thick disk. Later type K and M dwarfs from the thin disk populate the middle colour
range of the diagram (2\,$<$\,$I$--$K$\,$<$\,3).

We have selected new member candidates from the optical-infrared survey
by extracting objects falling to the right of the NextGen and Dusty
100\,Myr isochrones (Fig.\ \ref{phot_kik} and Fig.\ \ref{phot_kik_zoom}).
As a first step, we have extracted all objects falling
to the right of the NextGen 100\,Myr isochrones over
the whole magnitude range. The total number of newly
infrared-selected candidates is 103 out
of 22,129 detections in the 0.70 deg$^{2}$
surveyed area. Fig.\ \ref{phot_kik_zoom} provides a close-up
region of the ($I_{c}$--$K^{\prime}$,$K^{\prime}$) \CMD{}
where the new infrared-selected candidates are located.
Those new cluster member candidates are represented
by hexagons.
We have also overplotted the optically-selected
probable (filled triangles) cluster member candidates from
ByN02 for comparison.

We have cross-correlated the new infrared-selected candidates with
the recent release of the 2MASS database in order
to provide additional $J$ and $H$ magnitudes.
Analysis of the location of new candidates in various
colour-magnitude diagrams confirmed their status as new probable
members, except for AP416 which appears redder in the
optical (($R$--$I$)$_{c}$,$I_{c}$) \CMD{} (Fig.\ \ref{phot_rii}).
Further analysis of the membership status of the new
cluster candidates will be pursued in the next section.

As there is some evidence that low-mass brown dwarfs of age $\sim$\,100\,Myr
are better traced by the Dusty models (Jameson et al.\ \cite{jameson02}),
we have used those isochrones to identify a second set of low-mass
\APer{} candidate members. Based on the optical-to-infrared
colour-magnitude diagram (Figs.\ \ref{phot_kik} and \ref{phot_kik_zoom}),
we have extracted about 80 new objects. Five of them
remained brown dwarf candidates if they belong to the cluster;
the remainder being reddened background giants as discussed in
the next section. Two of the five candidates (AP399 and AP406)
were previously classified as probable brown dwarf members by ByN02\@.
The numbering of the new candidates follows from ByN02
one and starts at AP413\@.

%
%
%
\begin{table*}[!htbp]
\thispagestyle{headings}
\begin{center}
 \caption[New infrared-selected cluster candidates in \APer{}]{
New candidate members of the \APer{} cluster selected from the
optical-infrared survey using the NextGen and Dusty isochrones.
The names, coordinates in J2000, and $R_{c}$, $I_{c}$,
the 2MASS $J$, $H$, and $K_{s}$ magnitudes and $K^{\prime}$
magnitudes from our near-infrared observations are quoted.
The upper part of the
table provides the list of 18 optically-selected member candidates found
independently in the optical-to-infrared survey
whereas the middle and lower parts list the new infrared-selected probable
candidates. Among them, four new objects (AP430, AP431, AP432, and AP433)
are brown dwarf candidate members. The numbering of the new candidate
members follows the one described in ByN02 and starts at AP413\@.
}
 \normalsize
 \begin{tabular}{|l |c c | c | c | c | c | c | c |}
 \hline
Name & R.A. & Dec  & $R_{c}$ & $I_{c}$ & $J$ & $H$ & $Ks$  & $K^{\prime}$ \cr
 \hline
\multicolumn{9}{|c|}{Optically-selected cluster member candidates common to our survey}  \\
 \hline
AP329 & 03$:$23$:$56.36 & +48$:$09$:$21.1 & 17.16 & 15.48 & 13.88 & 13.31 & 13.04 & 12.95  \\ 
AP332 & 03$:$25$:$16.91 & +48$:$36$:$09.3 & 17.63 & 15.80 & 14.06 & 13.43 & 13.19 & 13.13  \\ 
AP334 & 03$:$22$:$45.51 & +48$:$21$:$33.4 & 17.82 & 15.96 & 14.25 & 13.66 & 13.34 & 13.35  \\ 
AP339 & 03$:$26$:$33.28 & +50$:$07$:$41.9 & 17.97 & 16.15 & 14.35 & 13.76 & 13.40 & 13.45  \\ 
AP343 & 03$:$23$:$48.49 & +48$:$36$:$43.3 & 18.01 & 16.21 & 14.62 & 14.07 & 13.72 & 13.66  \\ 
AP344 & 03$:$26$:$52.11 & +50$:$00$:$32.8 & 18.24 & 16.27 & 14.33 & 13.69 & 13.39 & 13.44  \\ 
AP309 & 03$:$22$:$40.69 & +48$:$00$:$33.8 & 18.23 & 16.38 & 14.49 & 13.88 & 13.58 & 13.59  \\ 
AP349 & 03$:$26$:$48.01 & +50$:$02$:$15.7 & 18.45 & 16.56 & 14.62 & 14.04 & 13.73 & 13.78  \\ 
AP353 & 03$:$24$:$48.69 & +48$:$49$:$47.3 & 18.78 & 16.68 & 14.61 & 14.00 & 13.65 & 13.63  \\ 
AP364 & 03$:$20$:$39.19 & +49$:$32$:$06.2 & 18.90 & 16.92 & 15.00 & 14.41 & 14.03 & 14.00  \\ 
AP365 & 03$:$28$:$22.98 & +49$:$11$:$24.3 & 19.06 & 17.03 & 15.02 & 14.33 & 14.15 & 14.20  \\ 
AP366 & 03$:$26$:$35.49 & +49$:$15$:$44.2 & 19.05 & 17.04 & 15.12 & 14.54 & 14.22 & 14.15  \\ 
AP369 & 03$:$26$:$45.11 & +50$:$25$:$06.7 & 19.46 & 17.26 & 14.88 & 14.26 & 13.88 & 13.84  \\ 
AP311 & 03$:$23$:$08.70 & +48$:$04$:$50.7 & 19.56 & 17.50 & 15.44 & 14.71 & 14.32 & 14.46  \\ 
AP378 & 03$:$27$:$01.00 & +49$:$14$:$41.2 & 19.71 & 17.67 & 15.61 & 14.94 & 14.50 & 14.62  \\ 
AP305 & 03$:$19$:$21.62 & +49$:$23$:$31.1 & 20.59 & 18.36 & 15.81 & 14.96 & 14.64 & 14.78  \\ 
AP399 & 03$:$25$:$48.55 & +50$:$01$:$00.8 & 22.71 & 20.15 &  ---  &  ---  &  ---  & 16.09  \\ 
AP406 & 03$:$23$:$09.86 & +48$:$16$:$30.0 & 23.60 & 20.78 &  ---  &  ---  &  ---  & 16.31  \\ 
 \hline
\multicolumn{9}{|c|}{New infrared-selected cluster member candidates (NextGen isochrones)}  \\
 \hline
AP413 & 03$:$24$:$17.75 & +48$:$07$:$36.0 & 17.01 & 15.50 & 14.03 & 13.34 & 13.05 & 13.06  \\ 
AP414 & 03$:$22$:$05.17 & +48$:$12$:$46.1 & 17.06 & 15.56 & 14.00 & 13.26 & 13.05 & 12.97  \\ 
AP415 & 03$:$23$:$59.79 & +48$:$03$:$57.6 & 17.29 & 15.96 & 14.48 & 13.82 & 13.62 & 13.44  \\ 
AP416 & 03$:$25$:$32.89 & +48$:$45$:$21.4 & 18.31 & 16.14 & 13.75 & 13.06 & 12.82 & 12.85  \\ 
AP417 & 03$:$28$:$16.33 & +50$:$05$:$51.6 & 17.91 & 16.34 & 14.57 & 13.91 & 13.63 & 13.63  \\ 
AP418 & 03$:$23$:$59.92 & +48$:$08$:$00.5 & 17.92 & 16.38 & 14.84 & 14.09 & 13.87 & 13.81  \\ 
AP419 & 03$:$22$:$40.50 & +48$:$19$:$35.0 & 18.09 & 16.40 & 14.80 & 14.09 & 13.83 & 13.78  \\ 
AP420 & 03$:$26$:$02.06 & +50$:$05$:$34.7 & 18.04 & 16.45 & 14.68 & 14.06 & 13.71 & 13.78  \\ 
AP421 & 03$:$23$:$18.80 & +48$:$04$:$25.4 & 18.61 & 16.77 & 14.75 & 14.07 & 13.77 & 13.87  \\ 
AP422 & 03$:$27$:$17.27 & +50$:$07$:$19.8 & 18.51 & 16.87 & 15.12 & 14.41 & 14.13 & 14.12  \\ 
AP423 & 03$:$22$:$09.83 & +48$:$16$:$43.8 & 18.53 & 16.93 & 15.30 & 14.66 & 14.28 & 14.19  \\ 
AP424 & 03$:$23$:$17.96 & +47$:$59$:$01.7 & 18.69 & 17.07 & 15.31 & 14.54 & 14.20 & 14.28  \\ 
AP425 & 03$:$27$:$25.41 & +50$:$05$:$05.4 & 19.12 & 17.46 & 15.61 & 14.79 & 14.56 & 14.51  \\ 
AP426 & 03$:$24$:$08.12 & +48$:$48$:$30.0 & 19.53 & 17.54 & 15.50 & 15.03 & 14.44 & 14.59  \\ 
AP427 & 03$:$23$:$04.86 & +48$:$16$:$11.3 & 19.68 & 17.66 & 14.61 & 15.15 & 14.61 & 14.62  \\ 
AP428 & 03$:$23$:$05.59 & +48$:$09$:$00.7 & 19.95 & 18.10 & 16.12 & 15.07 & 14.76 & 14.97  \\ 
AP429 & 03$:$27$:$07.06 & +50$:$09$:$22.7 & 20.15 & 18.12 & 15.75 & 15.00 & 14.77 & 14.78  \\ 
AP430 & 03$:$26$:$16.24 & +50$:$18$:$43.3 & 23.07 & 20.57 &  ---  &  ---  &  ---  & 16.03  \\ 
 \hline
\multicolumn{9}{|c|}{New infrared-selected cluster member candidates (Dusty isochrones)}  \\
 \hline
AP431 & 03$:$24$:$06.45 & +48$:$23$:$11.5 & 22.20 & 19.82 &  ---  &  ---  &  ---  & 16.07  \\ 
AP432 & 03$:$27$:$30.59 & +49$:$11$:$09.6 & 22.48 & 20.06 &  ---  &  ---  &  ---  & 16.27  \\ 
AP433 & 03$:$20$:$14.94 & +49$:$31$:$46.6 & 23.42 & 21.00 &  ---  &  ---  &  ---  & 16.39  \\ 
 \hline
\end{tabular}
 \label{new_cand_APer}
\end{center}
\end{table*}
%

%
%
\subsection{Colour-colour diagram}
\label{ccd}

\begin{figure}[htbp]
\begin{center}
\includegraphics[width=\columnwidth, angle=0]{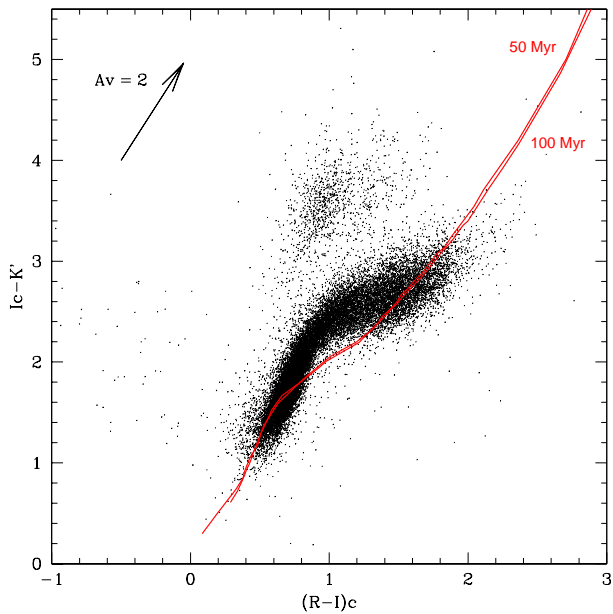}
\caption{
Colour-colour (I$_{c}$--$K^{\prime}$,($R$--$I$)$_{c}$).
Over 22,000 detections are plotted as black dots.
Overplotted are the NextGen isochrones from Baraffe et al.\ (\cite{baraffe98})
for 50 and 100 Myr, age bracketing the one estimated for $\alpha$\,Persei
assuming a distance of 182\,pc.
The giant branch (roughly at ($R$--$I$)$_{c}$\,$\sim$\,1.0 and $I$--$K'$\,$\sim$\,3.5)
is well populated in this diagram, showing that
\APer{} is located towards the galactic plane.
}
\label{phot_ikri}
\end{center}
\end{figure}

\begin{figure}[htbp]
\begin{center}
\includegraphics[width=\columnwidth, angle=0]{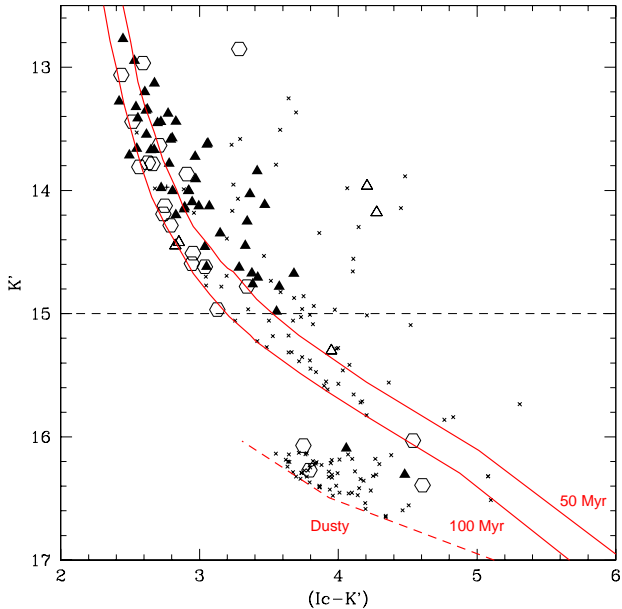}
\caption{
Zoom of the colour-magnitude ($I_{c}$--$K^{\prime}$,$K^{\prime}$)
where the new infrared-selected member candidates are located.
Overplotted are isochrones from Baraffe et al.\ (\cite{baraffe98})
for 50 and 100 Myr (solid lines),  and
the Dusty (dashed line; Chabrier et al.\ \cite{chabrier00}) 100\,Myr isochrones,
assuming a distance of 182\,pc for the cluster.
Filled triangles are probable cluster member
candidates extracted by the optical survey (ByN02) whereas open
traingles are possible members from the same study.
Open hexagons are new infrared-selected cluster member candidates
from our optical-infrared survey.
The diagonal crosses indicate the distant background giants.
The horizontal dashed line represents the stellar/substellar
boundary at M$_{K}$\,=\,8.3, assuming a distance modulus of 6.3
for the cluster and an age of 90\,Myr.
}
\label{phot_kik_zoom}
\end{center}
\end{figure}

\begin{figure}[htbp]
\begin{center}
\includegraphics[width=\columnwidth, angle=0]{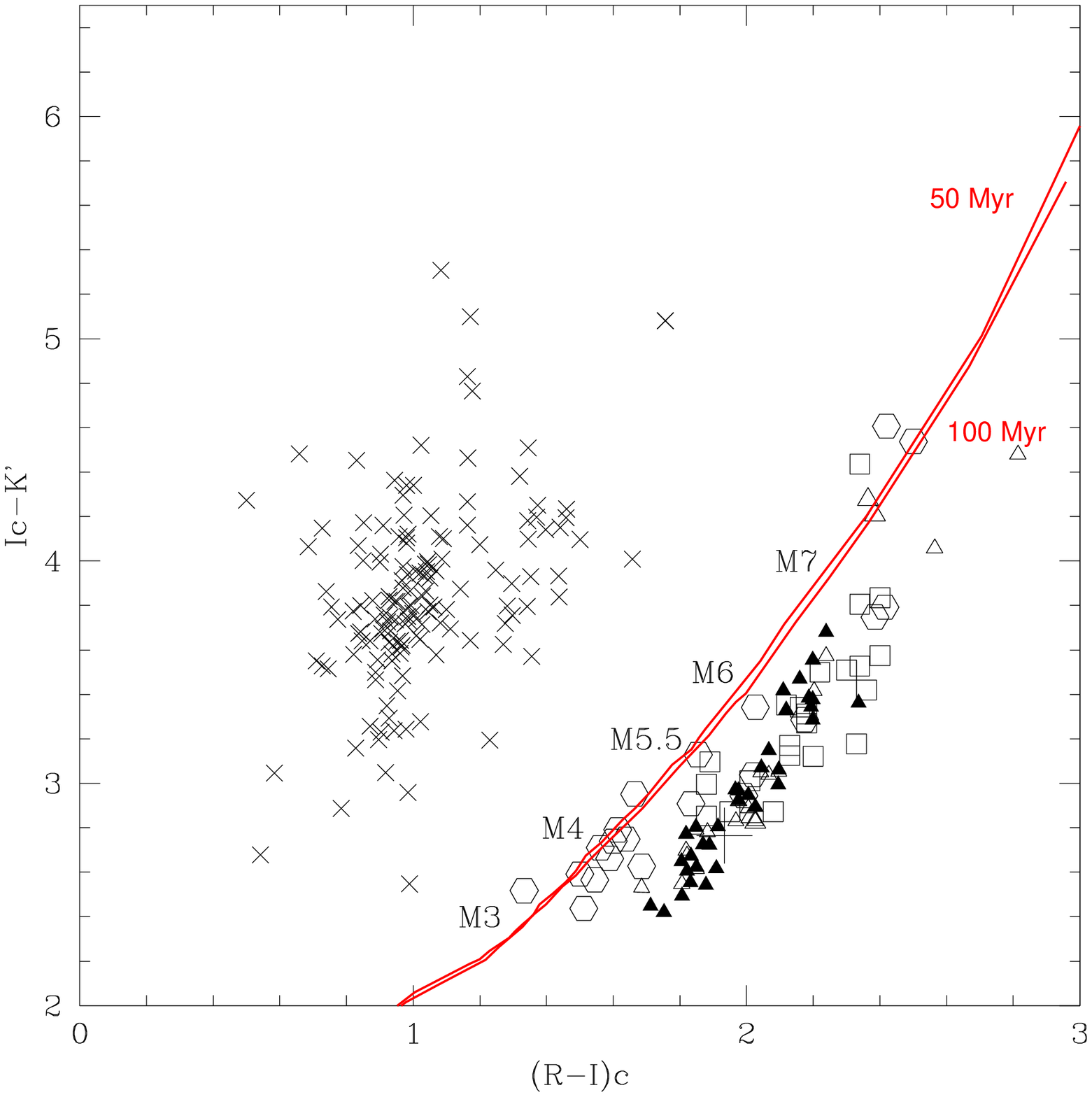}
\caption{
Colour-colour (I$_{c}$--$K^{\prime}$,($R$--$I$)$_{c}$).
Overplotted are NextGen isochrones from Baraffe et al.\ (\cite{baraffe98})
for 50 and 100 Myr, bracketing the age estimated for the cluster.
Filled and open triangles are probable and possible member
candidates extracted by the optical survey (ByN02).
Open squares are spectroscopically confirmed members of \APer{} (S99).
Open hexagons are new infrared-selected cluster candidates
from our optical-infrared survey. The diagonal crosses
indicate the distant background giants rejected as cluster members.
Spectral types for dwarfs from the young disk are
marked (Leggett \cite{leggett92}).
}
\label{phot_ikri_zoom}
\end{center}
\end{figure}

The \CCD{} is of prime importance to distinguish
probable cluster member candidates from distant background giants.
The photometry available in three broad-band filters
($R_{c}$, $I_{c}$, $K^{\prime}$) allows us to plot a colour-colour diagram
((($R$--$I$)$_{c}$,$I_{c}$--$K^{\prime}$) shown in Fig.\ \ref{phot_ikri}.
The giant branch is clearly visible and centred approximately on
($R$--$I$)$_{c}$\,$\sim$\,1 mag and $I_{c}$--$K^{\prime}$\,$\sim$\,3.5 mag.
A close-up of this diagram with the new infrared-selected cluster
member candidates is shown in Fig.\ \ref{phot_ikri_zoom}.
The symbols used in Fig.\ \ref{phot_kik_zoom} are identical
to the ones used in Fig.\ \ref{phot_ikri_zoom}.
The new candidates are listed in the middle and bottom panels
in Table \ref{new_cand_APer}.

%
%
%

%
%
%
%

%
%
%

Of the {{\bf103}} infrared-selected cluster member
candidates, two-thirds of them (65\,\%) turn out
to be reddened background giants (diagonal crosses on 
Fig.\ \ref{phot_ikri_zoom} and Fig.\ \ref{phot_kik_zoom})
rather than low-mass brown
dwarf candidates based on their location in the
((($R$--$I$)$_{c}$,$I_{c}$--$K^{\prime}$) colour-colour diagram.
The source of this contamination
lies in the fact that the $K^{\prime}$ band probes larger
distances through highly extincted regions close to the Galactic
plane; \APer{} lies at (l,b)\,=\,(146.5,--7.5).
The remaining infrared-selected candidates, which are located along the
Baraffe et al.\ (\cite{baraffe98}) NextGen isochrones in the 
colour-colour diagram, remain probable cluster member candidates.
However, half of these candidates (18 out of 36) were 
already extracted by
the optical survey, thus confirming their probable membership
(filled triangles in Fig.\ \ref{phot_ikri_zoom} and Fig.\ \ref{phot_kik_zoom};
upper part of Table \ref{new_cand_APer}).
The remaining 18 objects are new infrared-selected probable
member candidates not extracted as such by the optical survey
(open hexagons in Fig.\ \ref{phot_ikri_zoom} and Fig.\ \ref{phot_kik_zoom};
middle and lower part of Table \ref{new_cand_APer}).
One of these new member candidates (AP430)
has $R_c$, $I_c$, and $K'$ magnitudes
below the hydrogen-burning limit, indicating a
probable brown dwarf cluster member.
Another candidate (AP429) straddles the stellar/substellar
boundary and is possibly a brown dwarf. Finally,
AP428 lies close to the lithium depletion boundary
estimated by S99\@.

{\bf{Of the 80 cluster candidates selected to the right of the Dusty isochrones
(Fig.\ \ref{phot_kik_zoom}), five remain brown dwarf candidates if they belong
to the cluster. The other objects lie in the giant branch,
as seen in galactic models (Annie Robin, personal communication).
Thus, the contamination is very high in this part of the diagram,
reaching over 90\,\%}}

One can now ask the following question: why were those new
infrared-selected cluster member candidates missed in the optical selection?
To address this issue, we have compared the location of
the new candidates to previous members in the optical (($R$--$I$)$_c$,$I_c$)
\CMD{} (Fig.\ \ref{phot_rii}).

All the infrared-selected candidates located to the
right of the NextGen isochrones in the
($I_{c}$--$K^{\prime}$,$K^{\prime}$) \CMD{} also lie
to the right of the NextGen isochrones in the
(($R$--$I$)$_{c}$,$I_{c}$) \CMD{}.
The new infrared-selected candidates define a bluer sequence than
the optically-selected candidates by $\sim$\,0.2--0.3 mag in the
($I_{c}$,($R$--$I$)$_{c}$) colour-magnitude diagram (Fig.\ \ref{phot_rii})
and therefore remain likely new cluster member candidates.
Some objects follow the sequence defined by the probable
candidates extracted by ByN02\@.
One object, AP415, lies below the isochrones, suggesting that it is likely
a contaminant (Fig.\ \ref{phot_rii}).
It would therefore
be interesting to follow-up those new
probable candidates spectroscopically to determine the cut-off line of the
cluster sequence in the optical colour-magnitude diagram.
It could originate from the original candidate
selection procedure which made use of the Zero-Age-Main-Sequence
(Leggett et al.\ \cite{leggett92}) as a criterion (ByN02).
{\bf{The age of \APer{} is not questioned here 
since those objects are located on the
right of the 100\,Myr NextGen isochrone
(Baraffe et al.\ \cite{baraffe98}) in the (($R$--$I$)$_{c}$,$I_{c}$)
colour-magnitude diagram. The choice of the distance however affects
the number of selected candidates because larger/smaller distances to 
the cluster would move the isochrones in the colour-magnitude
diagrams. Choosing the HIPPARCOS distance of 190\,pc, for example, 
would lead to the identification of
two additional objects, AP418 and AP428, which are listed
in Table \ref{new_cand_APer}. The number of objects common
between the optical and the infrared selected samples did not vary,
and neither did the fractional contamination.}}

%
%
\section{Spectroscopic follow-up}
\label{specfollow}
%

%
%
\subsection{Observations}
\label{spec_obs}

Spectroscopic observations were carried out with the Twin
spectrograph mounted on the Calar Alto 3.5-m telescope in October/November
2002. Weather conditions were variable over the four night observing run.

The Twin CCD camera has a 2000\,$\times$\,800 pixel CCD detector with
a spectral resolution of 1.5 \AA{}/pixel. We have obtained moderate-resolution
(R\,$\sim$\,2000) optical spectra of 33 selected member candidates
using the red channel with the T07 grating covering 5800--8800\,\AA{}.
Two different slit widths, 1.5 and 2.1 arcsec, were used
according to the seeing conditions. Exposure times were adjusted
to take into account the brightness of the objects as well as
the weather conditions.

Our sample of objects can be divided into two categories.
First, a total of 29 optically-selected candidate members from ByN02 and S99
divided into four categories: 24 probable members, 1 possible member (AP350), 
and 4 non-members (AP327, AP336, AP338, and AP342).
Second, four new infrared-selected candidates reported in this paper
(AP414, AP416, AP417, and AP420)

Table \ref{spec_APer_table} lists all 33 objects observed
spectroscopically, their $I_{c}$ magnitudes,
the observing dates, and the exposure times as well as spectroscopic
results which are detailed in \S\ref{spectro_opt}.
Most of the objects were observed several times, each exposure
being shifted along the slit by $\sim$\,100 pixels.
Spectrophotometric standard stars Feige\,110 (Hamuy \cite{hamuy92})
and G191-B2B (Massey \& Gronwall \cite{massey90}) were observed twice
each during the night to flux calibrate the spectra.
Meanwhile, some template objects with known spectral types
were observed with the same set-up to derive accurate spectral type
classification. Dome flat fields, dark frames and He--Ar arc
lamps were taken before the beginning of each night.

%
%
%
\begin{sidewaystable*}[htbp]
\pagestyle{headings}
\begin{center}
 \caption[Spectroscopic data of \APer{} member candidates]{\small
                   List of all 33 objects observed with the red channel
                   of the T07 grating on the Twin spectrograph at the
                   Calar Alto 3.5-m telescope. Names, coordinates (in J2000),
                   $I_{c}$ magnitudes, observing date, exposure times,
                   equivalent widths for H$\alpha$ at 6563\,\AA{}
                   (negative values for emission lines),
                   equivalent widths {{\bf(in \AA{})}} for 
		   the K\,{\small{I}} and Na\,{\small{I}}
                   doublets at 7665/7699\,\AA{} and 8183/8195\,\AA{},
                   respectively, are included in the table.
                   Values obtained for three spectral indices,
                   TiO5 (Reid et al.\ \cite{reid95}),
                   VO-a (Kirkpatrick et al.\ \cite{kirkpatrick99}),
                   and PC3 (Mart\'{\i}n et al.\ \cite{martin99})
                   are given in columns 10, 11, and 12,
                   respectively. Former (ByN02) and new (this paper) membership
                   status as well as the spectral types (with an uncertainty
                   of half a subclass) are provided in columns 13 and 14\@.
                   The Y, Y$+$, Y?, and N letters stands for members, probable
                   members, possible members, and non-members, respectively.
                   The objects are ordered by increasing $I_{c}$ magnitude.
}
 \smallskip   
 \footnotesize
 \begin{tabular}{l cc  c  c  c  c  c  c  c  c  c  c  c}
 \hline
 \hline
\noalign{\smallskip}
Target  &       R.A. & Dec & $I_{c}$ & Date & ExpT & H$\alpha$ & K\,I  &  Na\,I  & Ti05 & VO-a & PC3 & Member & Sp.\@\,Type \cr
\noalign{\smallskip}
 \hline
\noalign{\smallskip}
AP327   & 03$:$20$:$31.74 & $+$49$:$39$:$59.6 &  15.06 & 01\,Nov\,02 & 3\,$\times$\,600\,s & ---  &  ---  &  {\bf{0.38}} & 0.985 (****) &  1.984 (****) &  1.109 (****) & N $\ra$ N & {\bf{early-M star}} \cr
AP329   & 03$:$23$:$56.34 & $+$48$:$09$:$21.0 &  15.48 & 31\,Oct\,02 & 1\,$\times$\,1800\,s & --6.5 & 2.0/1.3 & 2.1/2.4 & 0.390 (M4.0) & 2.001 (M4.8) & 1.236 (M4.7) & Y+ $\ra$ Y & M4.5\,$\pm$\,0.5 \cr
AP414   & 03$:$22$:$05.21 & $+$48$:$12$:$46.0 &  15.56 & 03\,Nov\,02 & 3\,$\times$\,600\,s & --- & 1.9/1.7 & 0.4/1.2 & 0.518 (M2.6) &  1.950 (M4.2) & 1.086 (M3.7) & N & M3.5\,$\pm$\,0.5 \cr
AP330   & 03$:$34$:$15.60 & $+$49$:$58$:$48.0 &  15.66 & 31\,Oct\,02 & 2\,$\times$\,1800\,s & --10.2 & 2.6/1.6 & 2.7/3.4 & 0.375 (M4.2) & 1.975 (M4.5) & 1.240 (M4.7) & Y+ $\ra$ Y & M4.5\,$\pm$\,0.5 \cr
AP283   & 03$:$24$:$38.78 & $+$48$:$17$:$17.1 &  15.70 & 10\,Oct\,02 & 2\,$\times$\,1200\,s & --4.2 & 2.3/1.6 & 2.4/2.8 & 0.386 (M4.0) & 1.987 (M4.6) & 1.232 (M4.7) & Y+ $\ra$ Y & M4.5\,$\pm$\,0.5 \cr
AP331   & 03$:$32$:$05.90 & $+$50$:$05$:$55.0 &  15.73 & 31\,Oct\,02 & 3\,$\times$\,900\,s & --7.9 & 2.2/2.6 & 2.7/1.8 & 0.341 (M4.5) & 1.973 (M4.5) & 1.299 (M5.1) & Y+ $\ra$ Y & M4.5\,$\pm$\,0.5 \cr
AP332   & 03$:$25$:$16.90 & $+$48$:$36$:$09.0 &  15.80 & 31\,Oct\,02 & 3\,$\times$\,800\,s & --8.2 & 2.8/1.9 & 3.3/2.5 & 0.319 (M4.8) & 1.992 (M4.7) & 1.307 (M5.2) & Y+ $\ra$ Y & M4.5\,$\pm$\,0.5 \cr
AP333   & 03$:$25$:$13.55 & $+$50$:$27$:$33.0 &  15.86 & 31\,Oct\,02 & 3\,$\times$\,800\,s & --5.7 & 3.2/2.7 & 2.7/3.4 & 0.380 (M4.1) & 1.942 (M4.1) & 1.316 (M5.2) & Y+ $\ra$ Y & M4.5\,$\pm$\,0.5 \cr
AP335   & 03$:$28$:$52.96 & $+$50$:$19$:$25.9 &  15.97 & 01\,Nov\,02 & 3\,$\times$\,1500\,s & --10.7 & 3.6/1.8 & 2.9/3.2 & 0.352 (M4.4) &  1.994 (M4.7) & 1.306 (M5.2) & Y+ $\ra$ Y & M4.5\,$\pm$\,0.5 \cr
AP336   & 03$:$31$:$55.30 & $+$49$:$08$:$31.0 &  16.04 & 01\,Nov\,02 & 3\,$\times$\,600\,s & ---  &  ---  &  {\bf{0.43}} & 0.972 (****) &  1.971 (****) &  1.088 (****) & N $\ra$ N & {\bf{early-M star}} \cr
AP337   & 03$:$34$:$07.50 & $+$48$:$32$:$08.0 &  16.07 & 01\,Nov\,02 & 3\,$\times$\,1200\,s & --7.9  & 2.5/2.3 & 2.5/2.3 & 0.295 (M5.0) & 2.006 (M4.8) & 1.341 (M5.4) & Y+ $\ra$ Y & M5.0\,$\pm$\,0.5 \cr
AP338   & 03$:$24$:$52.65 & $+$48$:$46$:$12.8 &  16.07 & 02\,Nov\,02 & 3\,$\times$\,600\,s & ---  &  ---  &  {\bf{1.0}}  & 0.964 (****) & 1.940 (****) & 0.953 (****) & N $\ra$ N & {\bf{early-M star}} \cr
{\bf{AP416}}   & 03$:$25$:$32.90 & $+$48$:$45$:$20.8 &  16.14 & 03\,Nov\,02 & 3\,$\times$\,750\,s & --- & 1.1/2.0 & 1.1/0.8 & 0.520 (M2.6) &  1.980 (M4.5) & 1.030 (M3.2) & N & M3.5\,$\pm$\,0.5 \cr
AP339   & 03$:$26$:$33.24 & $+$50$:$07$:$41.7 &  16.15 & 01\,Nov\,02 & 3\,$\times$\,1200\,s & --6.5  &  2.6/1.6 & 3.6/3.3 & 0.337 (M4.6) & 1.982 (M4.6) & 1.340 (M5.4) & Y+ $\ra$ Y & M5.0\,$\pm$\,0.5 \cr
AP340   & 03$:$21$:$34.85 & $+$48$:$16$:$28.7 &  16.16 & 01\,Nov\,02 & 3\,$\times$\,900\,s & --3.8  &  5.4/4.6 & 3.6/4.3 & 0.276 (M5.2) & 2.018 (M4.9) & 1.390 (M5.7) & Y+ $\ra$ Y & M5.0\,$\pm$\,0.5 \cr
AP341   & 03$:$31$:$03.39 & $+$50$:$24$:$41.6 &  16.16 & 01\,Nov\,02 & 3\,$\times$\,750\,s & --10.7 & 3.1/2.1 & 3.0/3.0 & 0.276 (M5.2) & 1.997 (M4.7) &  1.376 (M5.6) & Y+ $\ra$ Y & M5.0\,$\pm$\,0.5 \cr
AP344   & 03$:$26$:$52.00 & $+$50$:$00$:$33.0 &  16.27 & 01\,Nov\,02 & 3\,$\times$\,600\,s & --10.6 & 3.4/2.3 & 3.1/3.1 & 0.254 (M5.5) & 2.030 (M5.1) &  1.392 (M5.7) & Y+ $\ra$ Y & M5.5\,$\pm$\,0.5 \cr
AP345   & 03$:$33$:$45.80 & $+$50$:$08$:$53.0 &  16.30 & 01\,Nov\,02 & 3\,$\times$\,600\,s & --9.9 & 2.4/1.5 & 2.7/3.2 & 0.313 (M4.8) & 1.995 (M4.7) & 1.372 (M5.6) & Y+ $\ra$ Y & M5.0\,$\pm$\,0.5 \cr
AP346   & 03$:$21$:$30.06 & $+$48$:$49$:$23.2 &  16.32 & 01\,Nov\,02 & 3\,$\times$\,600\,s & --8.3 & 2.7/2.1 & 2.5/3.5 & 0.327 (M4.7) & 1.981 (M4.5) & 1.381 (M5.6) & Y+ $\ra$ Y & M5.0\,$\pm$\,0.5 \cr
{\bf{AP417}}   & 03$:$28$:$16.28 & $+$50$:$05$:$51.7 &  16.34 & 02\,Nov\,02 & 3\,$\times$\,1200\,s & --4.8 & 0.9/1.0 & 2.3/2.3 & 0.520 (M2.6) & 1.989 (M4.6) & 1.048 (M3.4) & N & M3.5\,$\pm$\,1.0 \cr
AP347   & 03$:$31$:$33.77 & $+$49$:$52$:$02.1 &  16.34 & 01\,Nov\,02 & 3\,$\times$\,600\,s & --6.9 & 3.1/1.9 & 3.0/3.0 & 0.288 (M5.1) & 1.962 (M4.4) & 1.368 (M5.6) & Y+ $\ra$ Y & M5.0\,$\pm$\,0.5 \cr
AP342   & 03$:$25$:$39.24 & $+$48$:$45$:$21.1 &  16.34 & 02\,Nov\,02 & 3\,$\times$\,600\,s & ---  &  ---  &  {\bf{0.4}}  & 0.958 (****) &  1.998 (****) &  1.043 (****) & N $\ra$ N & {\bf{early-M star}} \cr
AP309   & 03$:$22$:$40.65 & $+$48$:$00$:$33.6 &  16.38 & 01\,Nov\,02 & 3\,$\times$\,600\,s & --4.9 & 2.1/1.5 &  2.3/1.6  & 0.345 (M4.5) &  2.021 (M5.0) & 1.336 (M5.4) & Y+ $\ra$ Y & M5.0\,$\pm$\,0.5 \cr
{\bf{AP420}}   & 03$:$26$:$02.08 & $+$50$:$05$:$34.5 &  16.45 & 03\,Nov\,02 & 2\,$\times$\,750\,s & --- & --- & --- & 0.979 (****) &  1.970 (****) & 0.871 (****) & N & {\bf{G star}} \cr
AP349   & 03$:$26$:$47.90 & $+$50$:$02$:$16.0 &  16.56 & 02\,Nov\,02 & 3\,$\times$\,1200\,s & --9.6 & 2.4/2.2 &  3.3/3.4  & 0.293 (M5.0) &  1.986 (M4.6)& 1.373 (M5.6) & Y+ $\ra$ Y & M5.5\,$\pm$\,0.5 \cr
AP350   & 03$:$32$:$06.80 & $+$49$:$25$:$23.0 &  16.57 & 03\,Nov\,02 & 1\,$\times$\,750\,s & ---  &  ---  &  {\bf{1.0}}  & 1.017 (****) &  1.996 (****) &  0.913 (****) & Y? $\ra$ N & {\bf{early-M star}} \cr
AP351   & 03$:$28$:$47.80 & $+$50$:$02$:$01.0 &  16.64 & 02\,Nov\,02 & 3\,$\times$\,1200\,s & --7.6 & 2.5/2.1 & 3.5/3.7 & 0.295 (M5.0) & 2.041 (M5.2) & 1.322 (M5.3) & Y+ $\ra$ Y & M5.0\,$\pm$\,0.5 \cr
AP353   & 03$:$24$:$48.66 & $+$48$:$49$:$47.0 &  16.68 & 02\,Nov\,02 & 3\,$\times$\,1200\,s & --9.2 & 3.6/2.9 & 3.0/3.1 & 0.228 (M5.7) & 2.061 (M5.4) & 1.476 (M6.2) & Y+ $\ra$ Y & M6.0\,$\pm$\,0.5 \cr
AP354   & 03$:$27$:$31.64 & $+$48$:$53$:$23.3 &  16.69 & 02\,Nov\,02 & 3\,$\times$\,1200\,s & --11.1 & 3.4/2.6 & 2.9/3.2 & 0.224 (M5.8) & 2.069 (M5.5) & 1.458 (M6.1) & Y+ $\ra$ Y & M6.0\,$\pm$\,0.5 \cr
AP355   & 03$:$22$:$33.15 & $+$48$:$47$:$00.3 &  16.70 & 02\,Nov\,02 & 3\,$\times$\,1200\,s & --9.2 & 2.9/2.7 & 3.3/3.7 & 0.297 (M5.0) & 2.005 (M4.8) & 1.344 (M5.4) & Y+ $\ra$ Y & M5.0\,$\pm$\,0.5 \cr
AP363   & 03$:$24$:$00.33 & $+$47$:$55$:$29.7 &  16.88 & 02\,Nov\,02 & 3\,$\times$\,1200\,s & --14.9 & 3.1/2.0 & 3.7/4.1 & 0.288 (M5.1) & 1.992 (M4.7) &1.400 (M5.8) & Y+ $\ra$ Y & M5.5\,$\pm$\,0.5 \cr
AP364   & 03$:$20$:$39.16 & $+$49$:$32$:$06.0 &  16.92 & 02\,Nov\,02 & 3\,$\times$\,1200\,s & --8.2 & 2.4/2.8 & 3.9/4.3 & 0.430 (M3.6) & 2.028 (M5.0) & 1.390 (M5.7) & Y+ $\ra$ Y & M5.5\,$\pm$\,1.0 \cr
AP366   & 03$:$26$:$35.50 & $+$49$:$15$:$43.8 &  17.04 & 02\,Nov\,02 & 3\,$\times$\,1200\,s & --10.5 & 3.4/2.3 & 3.1/3.1 & 0.244 (M5.6) & 1.987 (M4.6) & 1.417 (M5.9) & Y+ $\ra$ Y & M6.0\,$\pm$\,0.5 \cr
 \hline
\end{tabular}
\label{spec_APer_table}
\end{center}
\end{sidewaystable*}
%

%
%
%
\begin{figure*}[!htbp]
\begin{center}
\includegraphics[width=\linewidth, angle=0]{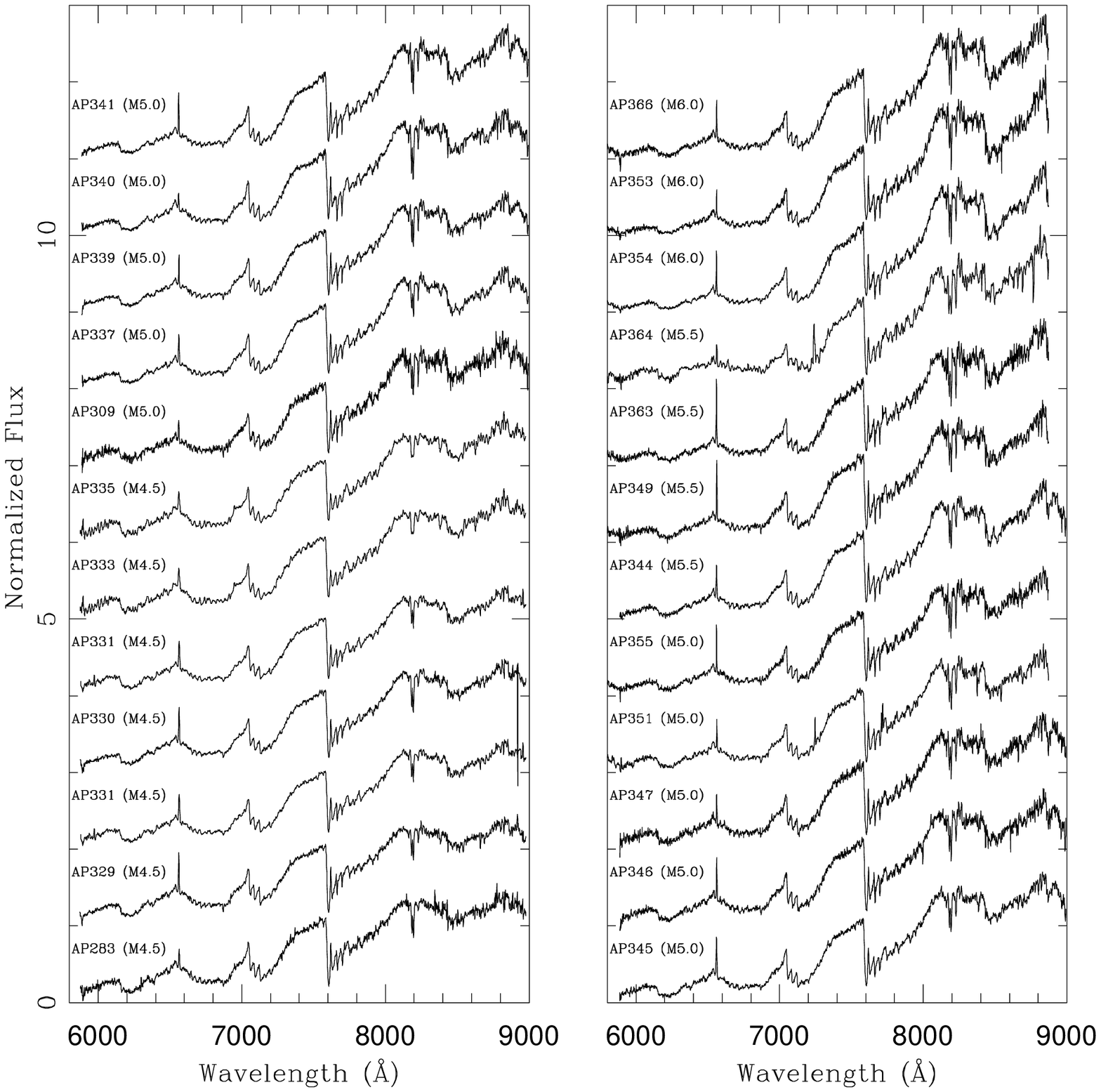}
\caption{
Spectra of 24 objects confirmed as cluster members via optical spectroscopy.
The spectral types quoted into brackets after the name
of the target range from M4.5 to M6.0, with a typical
uncertainty of order half a subclass.
Typical features of M dwarfs are clearly seen on the spectra,
including TiO and VO absorption broad bands as well as the
gravity-sensitive K\,{\small{I}} and Na\,{\small{I}}
doublets at 7665/7699\,\AA{} and 8183/8195\,\AA{}, respectively.
All targets exhibit H$\alpha$ in emission at 6563\,\AA{}.
An arbitrary constant has been added to each spectrum for clarity.
}
\label{spectro_APer}
\end{center}
\end{figure*}

%
%
\subsection{Data reduction}
\label{spec_datareduc}

The data reduction was standard and involved subtracting an
averaged bias frame and dividing
by a median-combined dome flat field. Wavelength calibration was made using He
and Ar lines throughout the whole wavelength range (RMS better than 0.4\,\AA{}). 
Flux calibration
was achieved using an averaged sensitivity function computed
with several exposures of the spectrophotometric standard stars.
If several spectra were obtained for one object,
the resulting co-added spectrum was the median of each individual 1-D spectrum
in order to remove the cosmic rays.
Optical spectra of the candidates confirmed as cluster members via 
spectroscopy are displayed in Fig.\ \ref{spectro_APer}.

\subsection{Spectroscopy of optical candidates}
\label{spectro_opt}

\subsubsection{Spectral classification}
\label{spectro_opt_class}

We have assigned a spectral type to each individual confirmed
member with an uncertainty of half a subclass, according to the M dwarf
classification schemes defined by Kirkpatrick et al.\ (\cite{kirkpatrick99})
and Mart\'{\i}n et al.\ (\cite{martin99}).
We have computed several spectral indices
including TiO5 (Reid et al \cite{reid95}),
VO-a (Kirkpatrick et al.\ \cite{kirkpatrick99}), and
the PC3  index (Mart\'{\i}n et al.\ \cite{martin99}).
The values of the three spectral indices, listed in Table \ref{spec_APer_table},
were generally consistent, although gravity might influence the 
computation of some spectral indices due to the young age
of $\alpha$\,Per compared with field dwarfs (Mart\'{\i}n et al.\ \cite{martin96}).

To derive a self-consistent classification and not
rely solely on spectral indices, we have compared each individual
target with spectra of template M dwarfs of similar spectral types, including
2MASS J2300189+121024, 2MASS J0244463+153531A\&B,
2MASS J0242252+134313 (Kirkpatrick et al.\ \cite{kirkpatrick99})
and 2MASS J0435490+153720 (Gizis et al.\ \cite{gizis99}), with
spectral types M4.5, M5.0, M5.5, M6.5, and M6.0, respectively.
In addition, due to possible differences in
telescope/instrument configurations
and detector sensitivities, other comparison objects with
known spectral types, including GJ251 (M3.0),
LHS1417 (M4.0), LHS0168 (M5.0), LHS1326 (M6.0),
and LHS0248 (M6.5) were observed with the same set-up as our targets.
The three different spectral type estimates yielded consistent
results with uncertainties of half a subclass and the
derived spectral types are consistent with the colours and membership
of the cluster.

%
%
%
\subsubsection{Chromospheric activity}
\label{spectro_opt_Halpha}

\begin{figure}[hptb]
\begin{center}
\includegraphics[width=\columnwidth, angle=0]{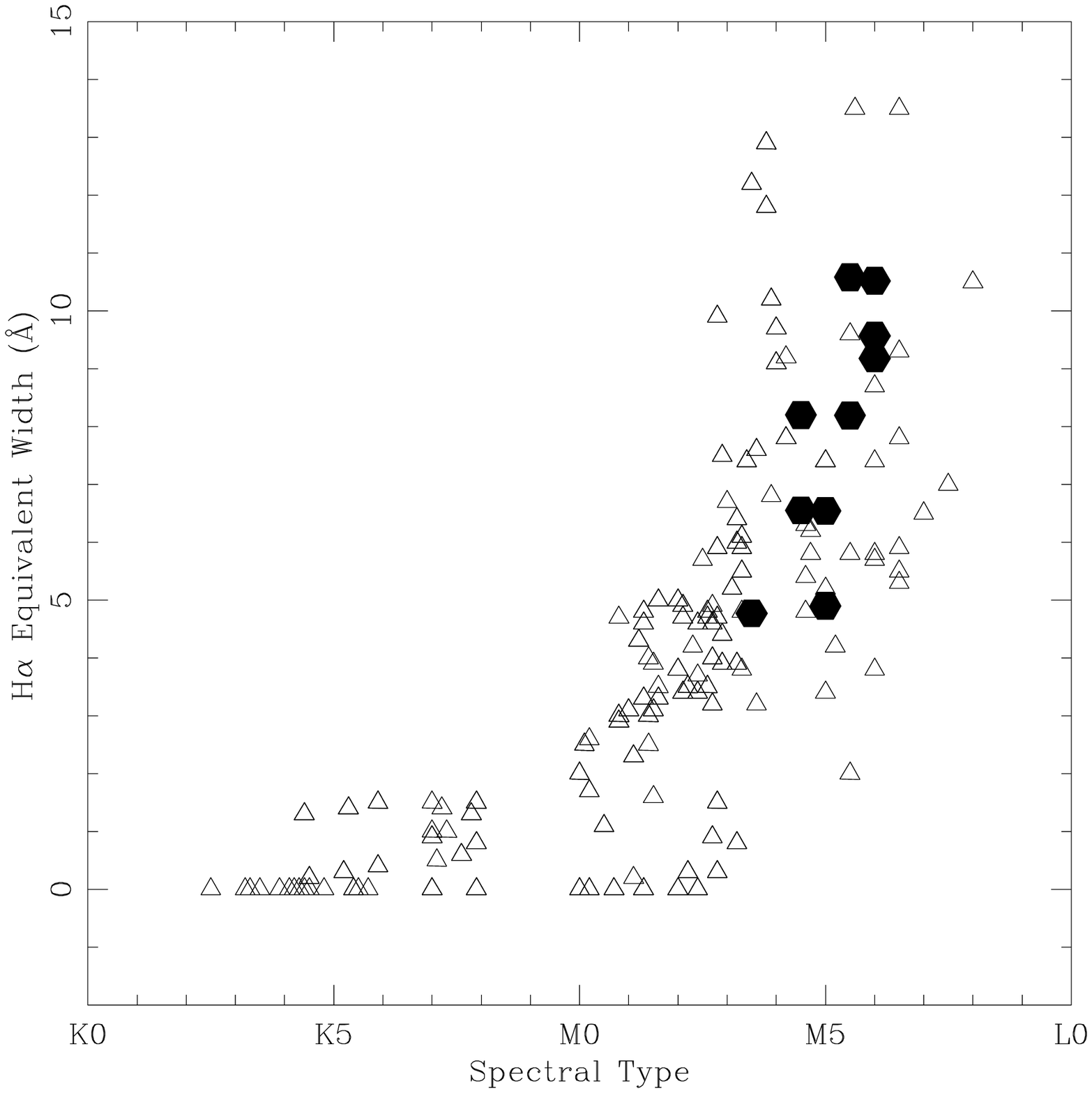}
\caption{
H$\alpha$ equivalent widths versus spectral type for all
spectroscopically confirmed cluster members in the \APer{} cluster.
Open triangles indicate members listed in Prosser (\cite{prosser92}, \cite{prosser94})
and S99\@. Our H$\alpha$ equivalent
widths are indicated with filled hexagons and are consistent
with measurements of cluster members.
}
\label{Halpha}
\end{center}
\end{figure}

As a sign of youth,
chromospheric activity is a useful criterion to ascertain the
membership of selected cluster candidates.
The H$\alpha$ emission line at 6563\,\AA{} is clearly detected 
in {\it{all}} 24 probable candidates from ByN02
(filled hexagons in Fig.\ \ref{Halpha}).
The H$\alpha$ equivalent widths range from 4 to 15\,\AA{} and
are consistent with previous measurements in \APer{} obtained by
Zapatero Osorio et al.\ (\cite{zapatero96}),
Prosser (\cite{prosser92}, \cite{prosser94}), and
S99 in the M3--M6 spectral type range
(open triangles in Fig.\ \ref{Halpha}).
However, our sample contains too few objects to probe the turnover
around M3--M4 in \APer{} (Zapatero Osorio et al.\ \cite{zapatero96})
caused by the transition
from radiative to convective cores occurring at 0.3--0.2\Msun{}, regardless
of the age (D'Antona \& Mazzitelli \cite{dantona94}).

%
%
%

Furthermore, the strength of the H$\alpha$ emission line
in the \APer{} member candidates is stronger than in field dwarfs
of similar spectral types, although the chromospheric activity
in field M dwarfs reaches a maximum around M6--M7
(Hawley et al.\ \cite{hawley96}) and can be as high as
in young magnetically active objects (Gizis et al.\ \cite{gizis02}).
For the maximum of activity in young clusters, see also
the compilation of the Pleiades, \APer{}, and IC2391 in
Barrado y Navascu\'es \& Mart\'{\i}n (\cite{barrado03}).

Moreover, H$\alpha$ equivalent widths measured
in M4--M8 cluster members in the Pleiades
are typically greater than 3\,\AA{}. Although arbitrary,
the 3\,\AA{} value reflects the lower envelope of equivalent
widths in the Pleiades (Hodgkin et al.\ \cite{hodgkin95}),
consistent with our measurements.
On the side of the high level of chromospheric activity, our
findings are in agreement with the saturation limits of the
H$\alpha$ emission line in young clusters as described by
Barrado y Navascu\'es \& Mart\'{\i}n (\cite{barrado03}).

Therefore, the detection of the H$\alpha$ emission line in all probable
cluster members adds support
to the belief that they are indeed members of $\alpha$\,Per.

\subsubsection{Surface gravity}
\label{spectro_opt_gravity}

To further constrain the membership of the photometric
candidates, we have computed the strengths of gravity-sensitive
features, including the K\,{\small{I}} and Na\,{\small{I}} doublets
at 7665/7699\,\AA{} and 8183/8195\,\AA{}, respectively 
(Table \ref{spec_APer_table}).
The gravity decreases with younger ages, implying that young
pre-main-sequence objects should harbour lower gravity and thus
smaller equivalent widths.

\begin{figure}[hptb]
\begin{center}
\includegraphics[width=\columnwidth, angle=0]{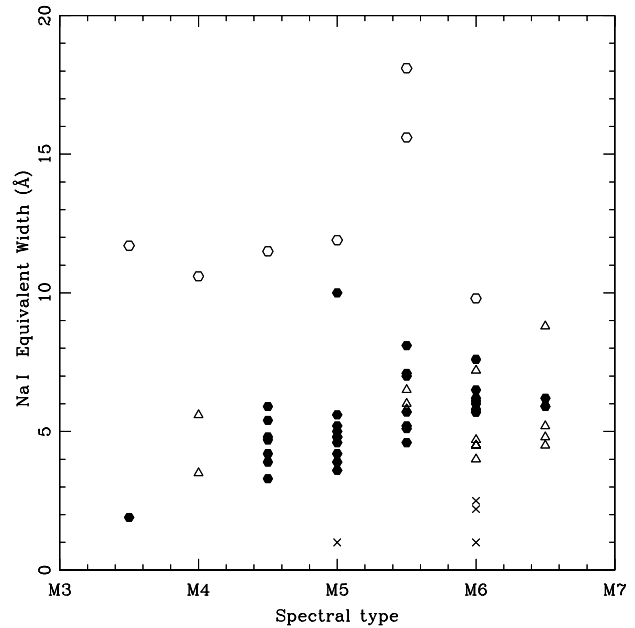}
\caption{
Na\,{\small{I}} equivalent widths (in \AA{}) versus spectral type for all
spectroscopically confirmed cluster members in the \APer{} cluster
(filled hexagons). Spectroscopic data for the Pleiades
(open triangles; Martin et al.\ \cite{martin96}),
$\sigma$\,Orionis (crosses; B\'ejar et al.\ \cite{bejar99}),
and field dwarfs (open circles; Mart\'{\i}n et al.\ \cite{martin96})
are overplotted for comparison.
Our measurements are compatible with those in the Pleiades,
suggesting that those candidates are indeed members of \APer{}.
}
\label{NaI} 
\end{center}
\end{figure}

The equivalent widths of the K\,{\small{I}} doublet in \APer{}
are slightly smaller than for field M dwarfs of similar spectral types
Similarly, a difference in the Na\,{\small{I}} doublet equivalent widths
between \APer{} and field dwarfs is observed, indicative of lower 
surface gravities (open hexagons in Fig.\ \ref{NaI}; 
Mart\'{\i}n et al.\ \cite{martin96}). 
The equivalent widths of the Na\,{\small{I}}
doublet are on the order of 3--7\,\AA{}, consistent with measurements
in the Pleiades (open triangles in Fig.\ \ref{NaI}; 
Mart\'{\i}n et al.\ \cite{martin96}) and larger
than those in $\sigma$\,Orionis (crosses in Fig.\ \ref{NaI}; 
B\'ejar et al.\ \cite{bejar99}).
The differences are nevertheless small, as predicted:
the gravity difference between 100\,Myr-old cluster members
and old (1--5\,Gyr) field dwarfs is less than 0.2 dex according to
the NextGen models (Baraffe et al.\ \cite{baraffe98}).

%
%
%
%

%
%
\subsection{Spectroscopy of infrared-selected candidates}

We have obtained optical spectroscopy of a subsample of four
infrared-selected member candidates in the \APer{} cluster (AP414, AP416,
AP417, and AP420) to test the optical-to-infrared selection
method presented in \S\ref{WFNIR}. We have classified those
objects according to their spectral types and chromospheric
activity as described below.

One object (AP420) is clearly a non-member of \APer{}
due to its spectra exhibiting H$\alpha$ and Na{\small {I}}D in
absorption.
We rejected two other candidates, AP414 and AP416, as possible
members due to the non-detection of the H$\alpha$ emission line
and colours inconsistent with membership. They are likely
contaminating field dwarfs with a spectral type of M3.5\,$\pm$\,0.5\@.
The small overlap available between the catalogue of
Deacon \& Hambly (\cite{deacon04}) and our survey shows that
both objects have a membership probability lower than 60\,\%.

The last object, AP417, also exhibits H$\alpha$ in emission.
While the equivalent width is consistent with previous studies of sources
in \APer{}, it is on the weak side. In addition, the sequence of $I_{c}$
magnitudes versus spectral type obtained for confirmed optically-selected
members would predict a spectral type of M5 for this object
(Table \ref{spec_APer_table}).
However, optical spectroscopy classifies it as an significantly
earlier M3.5\@. Therefore, we consider AP417 to be a non-member,
likely a field dwarf.

In summary, the optical-infrared method presented in this paper
to select cluster candidates in \APer{} is affected by field
dwarfs and reddened objects.

%
%
%
\begin{figure}[hptb]
\begin{center}
\includegraphics[width=\columnwidth, angle=0]{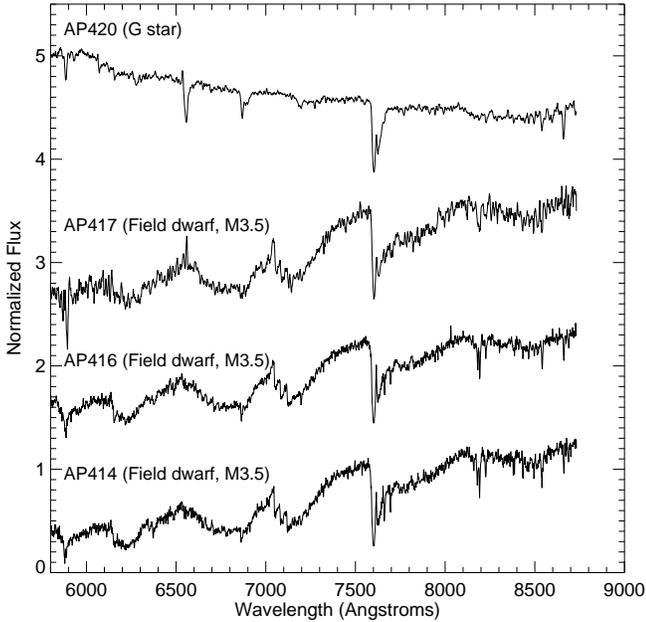}
\caption{
Optical spectra of the infrared-selected candidates rejected as cluster
members. From top to bottom are displayed the optical spectra of
AP414, AP416, AP417, and AP420\@.
}
\label{spec_APer_NM}
\end{center}
\end{figure}
%

%
%
\section{Discussion}
\label{spectro_discussion}

Our spectroscopic observations of the optically-selected candidates confirms
membership in each case. All probable members exhibit
H$\alpha$ in emission and have gravity measurements consistent with
young objects. Those members
span spectral types M4--M6, just above the M6.5 boundary which delineates
the stellar/substellar border in young open clusters (Luhman \cite{luhman99}).
Therefore, these are stellar components of the cluster with masses between
0.4 and 0.12\Msun{} according to the NextGen isochrones
(Baraffe et al.\ \cite{baraffe98}), assuming a distance of 182\,pc 
and an age of 90\,Myr for \APer{}.
One object, AP350, previously classified as a possible member by
ByN02, was definitely rejected by our optical spectroscopy.
Sources previously classified as non-members are, indeed,
non-members. Hence, the original optical survey and additional
near-infrared imaging follow-up appear to provide an
excellent discriminant between cluster members
and contaminants.

Conversely, selection based on the optical-infrared colour-magnitude
diagram alone is poorer, as the contamination among selected objects is 
higher. The large majority of candidates selected from the
($I_c$--$K'$,$K'$) \CMD{} are reddened background giants ($\sim$\,70\,\%).
Half of the new infrared-selected candidates were previously
considered as probable members while the rest await spectroscopic
follow-up. In summary, the contamination of our infrared-selected sample
lies in the range 70--85\,\%, values higher than the estimated
values from the optical survey (28--40\,\%; ByN02). In addition,
none of the new infrared-selected candidates remain bona-fide
cluster members after spectroscopic observations.
{\bf{One of these objects, AP416, was already suspected to be a
possible contaminant because it was located above the cluster
sequence in Fig.\ \ref{phot_rii}.}}

Thus, although we find the use of the $K^{\prime}$-band a valuable
tool to constrain optical selection, one should be cautious u\-sing the 
optical-infrared ($I_c$--$K'$,$K'$) \CMD{} as the sole criterion
to extract cluster members. To find very low-mass stars
and brown dwarfs, most of the surveys in the Pleiades and other 
open clusters are conducted in the optical ($R$, $I$, and $Z$)
to avoid large optical-to-infrared colours such as $I$--$K$,
where reddened field dwarfs and distant background giants
exhibit similar colours as young pre-main sequence stars.

This poses a warning for the next
generation of wide-field near-infrared cameras (e.\@g.\@ VISTA,
WFCAM on UKIRT, WIRCam on CFHT, and Omega2000 on the
Calar Alto 3.5-m telescope) being used to derive accurate mass functions
down into the substellar regime in young open clusters. As an example, 
the UKIRT Infrared Deep Sky Survey (UKIDSS; Hambly et al.\ \cite{hambly03}) 
Galactic Cluster Survey (GCS)
project aims to reach down to $\sim$\,25\,M$_{\rm Jup}$ using
$J$, $H$, and $K$ broad-band filters and to survey 50 times
more area in \APer{} than presented here. 
However, as we have seen the detection of low-mass stars and \BDs{} would be
optimised by including one or two optical filters (e.\@g.\@ $I$ and $z$)
in addition to the near-infrared filters.
Infrared surveys remain nevertheless very efficient in star-forming regions,
where the extinction often hampers the study of low-mass members.

%
\section{Conclusions}
\label{conclusion}

We have carried out a wide-field near-infrared
($K^{\prime}$-band) survey of a 0.70 deg$^{2}$ area in the
\APer{} cluster. Combining the new infrared photometry with
existing optical ($R_{c}$ and $I_{c}$) imaging, we have extracted
new objects based on their location in the
optical-infrared ($I_{c}$--$K^{\prime}$,$K^{\prime}$)
\CMD{}. However, the position of these 
candidates --- optical-infrared --- selected in
the \CCD{} revealed that $\sim$\,70\,\% of them
are contaminants, including
a large number of background giants due to the low
galactic latitude of the cluster. Half of the remaining candidates were
already identified by the optical survey of
ByN02; the rest are new member candidates, including four new \BDs{}. 
The optical (($R$--$I$)$_{c}$,$I_{c}$) \CMD{} shows that the new infrared-selected
candidates define a bluer sequence than previous probable members
in \APer{}.

We have presented moderate-resolution (R\,$\sim$\,2000) optical
(5800--8800\,\AA{}) spectroscopy of 29 optically-selected 
candidates identified by ByN02\@.
All probable members have spectral types, chromospheric activity,
and surface gravity measurements consistent with cluster
membership. They span $I_{c}$\,=\,15--17 mag, corresponding
to masses from 0.40 to 0.12\Msun{} in \APer{}, according to evolutionary
tracks. Thus, the cluster mass spectrum derived in this mass range by
ByN02 (dN/dM\,$\propto$\,M$^{-0.59\pm0.05}$) is now spectroscopically
confirmed. Other candidates (possible members and likely non-members)
were rejected as members on the basis of their spectra.

Finally, optical
spectroscopy of four infrared-selected candidate members
revealed spectral types and H$\alpha$ equivalent
widths inconsistent with cluster membership.
About 40 new infrared-selected candidates remain as possible
cluster members, including 18 objects already
classified as such by ByN02\@.
Their optical and optical-infrared colours are
consistent with membership but spectroscopic confirmation
is lacking.

%
%
%
\begin{acknowledgements}
This work was supported by the European Commission FP5 Research Training Network
``The Formation and Evolution of Young Stellar Clusters''
(HPRN-CT-2000-00155) and the DFG (CA239/3-1). 
We are grateful to Thomas Stanke who carried out infrared observations
of the cluster in December 1998\@.
NL thanks Richard Jameson fur useful discussion on the classification
of field stars.
NL thanks the Calar Alto observatory staff and Gyula Szokoly
for support and help during his first observing run at Calar Alto.
DByN is indebted to the Spanish ``Programa Ram\'on y Cajal''
and  AYA2001-1124-CO2 programs.
\end{acknowledgements}


%
%

\Online
\appendix

%
%
%
%
%
\begin{table*}[!h]
\begin{center}
 \caption[]{
Optical-infrared catalogue of all sources detected in the 0.70
square degree surveyed area in \APer{}. The total number of sources
with optical and infrared photometry is 22,129\@. Optical and infrared coordinates
(in J2000) are listed in columns 2--5\@. Magnitudes and their
associated errors in $R_{c}$, $I_{c}$, and $K^{\prime}$
are given in columns 6--8\@. The last column gives the
name of the object according to IAU convention.
}
 \label{IR_detection}
 \footnotesize
 \begin{tabular}{@{\hspace{1mm}}r@{\hspace{3mm}}c@{\hspace{3mm}}c@{\hspace{3mm}}c@{\hspace{3mm}}c@{\hspace{3mm}}c@{\hspace{3mm}}c@{\hspace{3mm}}c@{\hspace{3mm}}c@{\hspace{1mm}}}
 \hline
 \hline
Number & R.A.\,(Opt) & Dec\,(Opt) & R.A.\,(NIR) & Dec\,(NIR) & $R_{c}$  & $I_{c}$  & K$^{\prime}$  &  IAU Name \cr
 \hline
    1 & 03$:$28$:$28.13 & +49$:$36$:$05.5 & 03$:$28$:$28.16 & +49$:$36$:$05.5 & 19.746$\pm$0.008 & 18.481$\pm$0.007 & 16.134$\pm$0.020 & AP J032828+493605 \\
    2 & 03$:$28$:$28.10 & +49$:$35$:$46.3 & 03$:$28$:$28.13 & +49$:$35$:$46.2 & 20.814$\pm$0.008 & 19.316$\pm$0.007 & 16.700$\pm$0.033 & AP J032828+493546 \\
\ldots & \ldots & \ldots & \ldots & \ldots & \ldots & \ldots & \ldots & \ldots \cr
22129 & 03$:$21$:$58.24 & +47$:$58$:$28.5 & 03$:$21$:$58.21 & +47$:$58$:$28.4 & 21.458$\pm$0.015 & 19.920$\pm$0.013 & 17.394$\pm$0.052 & AP J032158+475828 \\
\noalign{\smallskip}
 \hline
 \end{tabular}
\end{center}
\end{table*}

\end{document}